\documentclass[prd,superscriptaddress,preprintnumbers,amsmath,nofootinbib,preprint]{revtex4}
\usepackage{amsmath, amssymb, graphics,epsfig}

\usepackage{slashed}

\usepackage{hyperref}

\def\bfnabla{\mbox{\boldmath $\nabla$}}

\def\bfsigma{\mbox{\boldmath $\sigma$}}
\def\lQ{\Lambda_{\rm QCD}}
\newcommand{\nn}{\nonumber}
\newcommand{\be}{\begin{equation}}
\newcommand{\ee}{\end{equation}}
\newcommand{\bea}{\begin{eqnarray}}
\newcommand{\eea}{\end{eqnarray}}
\def\al{\alpha}

\def\siml{{\
    \lower-1.2pt\vbox{\hbox{\rlap{$<$}\lower6pt\vbox{\hbox{$\sim$}}}}\ }}

\def\dsl{\,\raise.15ex\hbox{/}\mkern-13.5mu D}

\def\bfnabla{\mbox{\boldmath $\nabla$}}
\def\bfsigma{\mbox{\boldmath $\sigma$}}

\def\lQ{\Lambda_{\rm QCD}}

\newcommand{\eps}{\epsilon}

\newcommand{\cf}{C_F}

\newcommand{\Appendix}[1]%
    {%
     \section{#1}%
      }

\begin{document}

\title{HQET renormalization group improved Lagrangian at $\mathcal{O}(1/m^3)$ with leading logarithmic accuracy: Spin-dependent case}
\author{Xabier Lobregat}
\email{lobregat@ifae.es }
\author{Daniel Moreno}
\email{dmoreno@ifae.es}
\address{Grup de F\'\i sica Te\`orica, Dept. F\'\i sica and IFAE-BIST, Universitat Aut\`onoma de Barcelona,\\ 
E-08193 Bellaterra (Barcelona), Spain}
\author{Rudin Petrossian-Byrne}
\email{rudin.petrossian-byrne@physics.ox.ac.uk}
\address{Rudolf Peierls Centre for Theoretical Physics, University of Oxford,
1 Keble Road, Oxford OX1 3NP, United Kingdom}
\date{\today}

%%%%%%%%%%%%%%%%%%%%%%%%%%%%%%%%%%%%%%%%%%%%%%%%%%%%%%%%%%%%
\begin{abstract}
We obtain the renormalization group improved expressions of the Wilson coefficients associated to the  ${\cal O}(1/m^3)$ spin-dependent heavy quark 
effective theory Lagrangian operators, with leading logarithmic approximation, in the case of zero light quarks. We have employed the Coulomb gauge. 
\end{abstract}
\maketitle

\section{Introduction}

The expansion in inverse powers of the heavy quark mass is a useful tool for the study of hadrons containing heavy quarks. This expansion can be formulated more systematically in terms of an effective theory
described by an effective Lagrangian. For the one-heavy quark sector this
theory is heavy quark effective theory (HQET) \cite{HQET}. The HQET Lagrangian is also a key object in the description of systems with more than one heavy quark, in particular
in the description of the heavy quark-antiquark sector (i.e. heavy quarkonium), since the HQET Lagrangian corresponds to one of the building blocks of the 
nonrelativistic QCD (NRQCD) Lagrangian~\cite{NRQCD,Bodwin:1994jh}. The Wilson coefficients of the HQET Lagrangian also enter into 
the Wilson coefficients of the operators (i.e. the potentials) of the pNRQCD Lagrangian \cite{Pineda:1997bj,Brambilla:1999xf}, 
an effective field theory optimised for the description of heavy quarkonium systems near threshold (for reviews 
see Refs. \cite{Brambilla:2004jw,Pineda:2011dg}). The Wilson coefficients we compute in this paper 
are necessary ingredients to obtain the pNRQCD Lagrangian with next-to-next-to-next-to-leading logarithmic (NNNLL) accuracy. It should be noted that the Wilson coefficients computed in this paper are not necessary to obtain the 
complete heavy quarkonium spectrum with NNNLL accuracy, nor the production and annihilation of heavy quarkonium with NNLL 
precision, unlike their spin-independent counterparts (see Ref. \cite{rgsddelta,inpreparation}), but they may become important at the next order. At least, 
they must be studied. These results are also instrumental in the determination of higher order
logarithms for NRQED bound states, such as hydrogen and muonic hydrogen-like atoms.

At present, the operator structure of the HQET Lagrangian, and the tree-level values of their Wilson coefficients, is known to ${\cal O}(1/m^3)$ in 
the case without massless  quarks \cite{Manohar:1997qy}. The inclusion of massless quarks has been considered in Ref. \cite{Balzereit:1998am}.
The Wilson coefficients with leading logarithmic (LL) accuracy were computed in Refs. \cite{Finkemeier:1996uu,Blok:1996iz,Bauer:1997gs}
to ${\cal O}(1/m^2)$ and at next-to-leading order (NLO) in Ref. \cite{Manohar:1997qy} to ${\cal O}(1/m^2)$ (without dimension 6 heavy-light operators). The LL 
running to ${\cal O}(1/m^3)$ without the inclusion of spectator quarks was considered in Refs. \cite{Balzereit:1998jb,Balzereit:1998vh}, which turned out 
to have internal discrepancies between their explicit single log results and their own anomalous dimension matrix. The computation was reconsidered in 
Ref. \cite{chromopolarizabilities}, where the spin-independent results were corrected. This work is a follow-up to 
Ref. \cite{chromopolarizabilities} and, for this reason, is structured very similarly. Here we focus on spin-dependent operators and obtain the renormalization group improved Wilson coefficients of the HQET Lagrangian with LL 
accuracy to $\mathcal{O}(1/m^3)$. We do not include light quarks in our analysis.

The paper is divided as follows. In Sec. \ref{Sec:Lag} we introduce the HQET Lagrangian. Sec. \ref{Sec:Compton} is dedicated to the study of 
the spin-dependent part of Compton scattering, performed in order to find physical combinations of Wilson coefficients. In Sec. \ref{Sec:RGE} we find the 
renormalization group equations (RGE) for these Wilson coefficients. The QCD case is considered in Sec. \ref{sec:RGeq}, and the QED case 
in Sec. \ref{sec:RGeqqed}. The solution of these equations is studied in Sec. \ref{Sec:numerics}. In Sec \ref{Sec:Comparison} we perform a detailed 
comparison between our results and the ones found in Refs. \cite{Balzereit:1998jb,Balzereit:1998vh}. Our conclusions are summarized in Sec. \ref{sec:conclusions}. Finally, in 
App. \ref{sec:fr} we present some new Feynman rules needed for the computation.

\section{HQET Lagrangian without light fermions}
\label{Sec:Lag}

The HQET Lagrangian is defined uniquely up to field redefinitions.
In this paper we use the HQET Lagrangian density for a quark of mass $ m \gg \lQ$, in the special frame $v=(1,0,0,0)$ given in 
Ref. \cite{Manohar:1997qy}:
\bea
&& 
{\cal L}_{\rm HQET}={\cal L}_g+{\cal L}_{Q}
%+{\cal L}_{\chi_{c}}+{\cal L}_{\psi\chi_{c}}
\,,
\label{LagHQET}
\\
\nn
\\
&&
{\cal L}_g=-\frac{1}{4}G^{\mu\nu \, a}G_{\mu \nu}^a +
c_1^{g}\frac{g}{4m^2} 
 f^{abc} G_{\mu\nu}^a G^{\mu \, b}{}_\alpha G^{\nu\alpha\, c}+
{\cal O}\left(\frac{1}{m^4}\right),
\label{Lg}
\\
\nn
\\
&&
\nn
{\cal L}_{Q}=
Q^{\dagger} \Biggl\{ i D_0 + \frac{c_k}{ 2 m} {\bf D}^2 
+ \frac{c_F }{ 2 m} {\bfsigma \cdot g{\bf B}} 
\\
&& \nn
+\frac { c_D}{ 8 m^2} \left({\bf D} \cdot g{\bf E} - g{\bf E} \cdot {\bf D} \right) + i \, \frac{ c_S}{ 8 m^2} 
{\bfsigma \cdot \left({\bf D} \times g{\bf E} -g{\bf E} \times {\bf D}\right) }
\\
&& \nn
+\frac {c_4 }{ 8 m^3} {\bf D}^4 + i c_M\, g { {\bf D\cdot \left[D \times B
\right] + \left[D \times B \right]\cdot D} \over 8 m^3}
%\nonumber\\
%&&
+ c_{A_1}\, {g^2}\, {{\bf B}^2-{\bf E}^2 \over 8 m^3}- 
c_{A_2} { {g^2}{\bf E}^2 \over 16 m^3} 
\\ 
&& + c_{W_1}\, g { \left\{ {\bf D^2,\bfsigma
\cdot B }\right\}\over 8 m^3} - c_{W_2}\, g { {\bf D}^i\, {\bf \bfsigma
\cdot B} \, {\bf D}^i  \over 4 m^3}+ c_{p'p}\, g { {\bf \bfsigma \cdot D\, B \cdot D + D
\cdot B\, \bfsigma \cdot D}\over 8 m^3} 
\nn
\\
&&
+ c_{A_3}\, {g^2}\, \frac{1}{N_c}{\rm Tr}\left({{\bf B}^2-{\bf E}^2 \over 8 m^3}\right)- 
c_{A_4}\, {g^2}\, \frac{1}{N_c}{\rm Tr}\left({{\bf E}^2 \over 16 m^3}\right) 
\nn
\\
&&
+ i c_{B_1}\, g^2\, { {\bf \bfsigma \cdot \left(B
\times B - E \times E \right)} \over 8 m^3}
- i c_{B_2}\, g^2\, { {\bf \bfsigma \cdot \left( E \times E \right)} \over 8 m^3}
\Biggr\} Q+
{\cal O}\left(\frac{1}{m^4}\right)
%\delta {\cal L}^{(4)}_{\psi}
%,
%\\
%\label{Lchic}
%&& {\cal L}_{\chi_{c}} = {\cal L}_{\psi_1}(\psi_1 \rightarrow \chi_{2c}, g \rightarrow -g, T^a \rightarrow (T^a)^T, 
%m\rightarrow m_2,(1) \rightarrow (2)),
%\\
%&&
%{\cal L}_{\psi\chi_{c}} =
% - \frac{d_{ss}}{ mm_2} \psi_1^{\dag} \psi_1 \chi_{2c}^{\dag} \chi_{2c}
%+
% \frac{d_{sv} }{ mm_2} \psi_1^{\dag} {\bfsigma} \psi_1
%                         \chi_{2c}^{\dag} {\bfsigma} \chi_{2c}
%\nn
%\\
%&&
%-
 %\frac {d_{vs} }{ mm_2} \psi_1^{\dag} {\rm T}^a \psi_1
%                         \chi_{2c}^{\dag} ({\rm T}^a)^T \chi_{2c}
%+
 % \frac{d_{vv}}{ mm_2} \psi_1^{\dag} {\rm T}^a {\bfsigma} \psi_1
%                         \chi_{2c}^{\dag} ({\rm T}^a)^T {\bfsigma} \chi_{2c}
\,.
\label{Lhh}
\eea
Where $Q$ is the NR fermion field represented by a Pauli spinor. The components of the vector $\bfsigma$ are the Pauli matrices. We define
$i D^0=i\partial^0 -gA^0$, $i{\bf D}=i\bfnabla+g{\bf A}$,
${\bf E}^i = G^{i0}$ and ${\bf B}^i = -\eps^{ijk}G^{jk}/2$, where $\eps^{ijk}$ is
the three-dimensional totally antisymmetric tensor\footnote{
In dimensional regularization several prescriptions are possible for the $\eps^{ijk}$ tensors and $\bfsigma$, and the same prescription 
as for the calculation of the Wilson coefficients must be used.}
with $\eps^{123}=1$ and $({\bf a} \times {\bf b})^i \equiv \eps^{ijk} {\bf a}^j {\bf b}^k$. Note also that we have rescaled by a factor $1/N_c$ the 
coefficients $c_{A_{3,4}}$ following Ref. \cite{chromopolarizabilities}, as compared to the definitions given in Ref. \cite{Manohar:1997qy}.

\section{Compton scattering}
\label{Sec:Compton}

Ref. \cite{chromopolarizabilities} showed that it is possible for the Wilson coefficients 
associated to $1/m^3$ operators to be gauge dependent. For example, this is the case for $c_{A_2}$, which always appears in physical observables along
with $c_M$ (well-known to be gauge dependent because it is related with $c_D$ through reparametrization invariance, Ref. \cite{Manohar:1997qy}) in such a way 
that the combination
is gauge independent/physical. In order to explore the existence of other physical combinations involving the Wilson coefficients that we aim to 
calculate , i.e. $c_{W_1}$, $c_{W_2}$, $c_{p'p}$, $c_{B_1}$ and $c_{B_2}$, we compute the amplitude for Compton scattering of a heavy quark with a gluon $Qg\rightarrow Qg$.
In this section, we restrict to the the spin-dependent part of this process in HQET. We compute it at tree level up to $\mathcal{O}(1/m^3)$ in the mass expansion and in the Coulomb gauge 
(though obviously the amplitude for Compton scattering is a gauge independent quantity). 
We take incoming and 
outgoing quarks to have four-momentum $p=(E_1,\bf p)$ and $p'=(E_1',\bf p\,')$. We take gluon four-momenta as outgoing and label them
by $k_1$, $i$, $a$ and $k_2$, $j$, $b$ with respect to color and vector indices. This also implies the on-shell condition 
$k_1^0= -|\bf k_1|$ and $k_2^0= |\bf k_2|$. We work in the incoming quark rest frame, i.e $E_1=0$ and $\bf p=0$, so $\bf p\,'=-(\bf k_1 + \bf k_2)$ 
and $E_1'=-(k_1^0 + k_2^0)$. In addition, we define the unit vectors $\bf n_1=\bf k_1/|\bf k_1|$ and 
$\bf n_2=\bf k_2/|\bf k_2|$. The relation
\begin{equation}
 |{\bf k}_2|=\frac{|{\bf k}_1|}{1+\frac{|{\bf k}_1|}{m}(1 + {\bf n}_1\cdot{\bf n}_2)}
\end{equation}
holds from four-momenta conservation.

\begin{figure}[!htb]
	%\begin{center}    
	\includegraphics[width=0.80\textwidth]{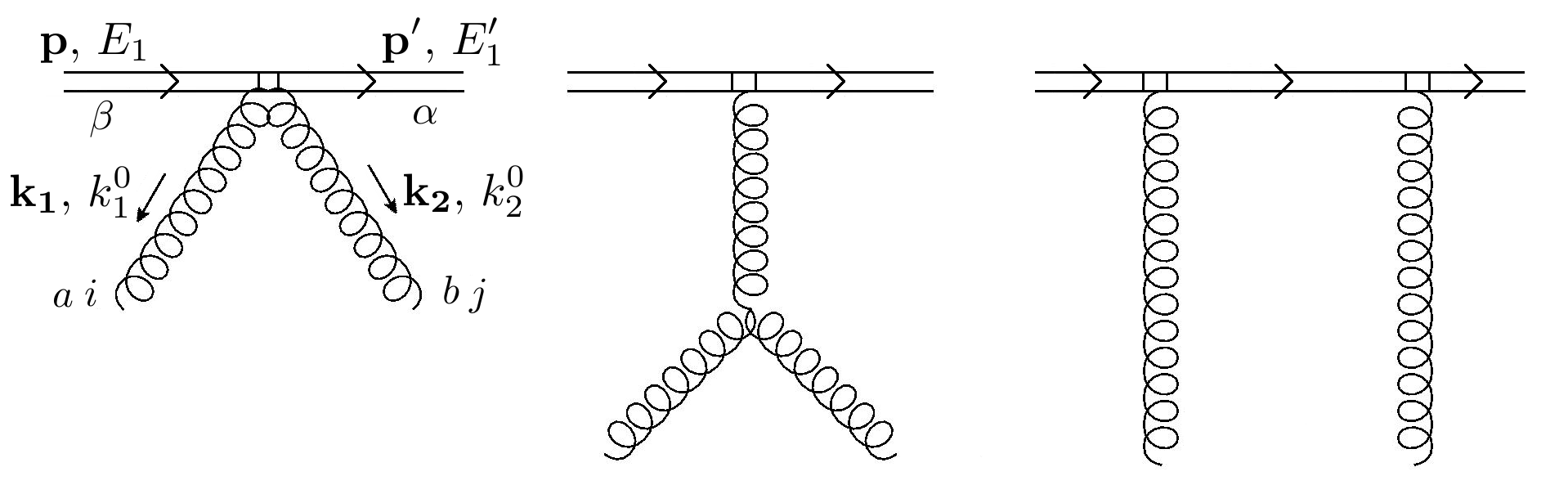} 
	%\includegraphics[width=0.35\textwidth]{Compton2.png}  
	%\includegraphics[width=0.17\textwidth]{Compton1.png}
	%\put(-45,-10){$c_{A_1}$} 
	%\put(-45,-10){$c_{A_1}$} 
	%\includegraphics[width=0.22\textwidth]{Compton3.png}
	%\put(-45,-10){$c_{A_1}$} 
	%
	%
	%\includegraphics[width=0.2\textwidth]{}
	%
\caption{Topologies contributing to Compton scattering at tree level up to ${\cal O}(1/m^3)$. Diagrams are generated from these topologies 
by considering all possible vertices contributing up to ${\cal O}(1/m^3)$.
\label{PlotsCompton}}   
%\end{center}
\end{figure}

By inserting the appropiate Wilson coefficients up to ${\cal O}(1/m^3)$, the topologies of the diagrams we have to consider for this computation are listed in Fig.~\ref{PlotsCompton}. The amplitude reads:

$$\mathcal{A}^{ij\,ab} = (c_{B_1}-2c_{W_1}- c_F^2 c_k  -c_S c_k)\frac{g^2}{16m^3}|{\bf k}_1|^2
( (\bfsigma\cdot{\bf n}_1){\bf n}_2^k \epsilon^{ijk}
 + (\bfsigma\cdot{\bf n}_2){\bf n}_1^k \epsilon^{ijk} $$
$$+ \bfsigma^i ({\bf n}_1\times{\bf n}_2)^j 
+ \bfsigma^j ({\bf n}_1\times{\bf n}_2)^i] [T^a,T^b]_{\alpha\beta}$$
$$-  (2c_{W_1} - 2c_{W_2}  +2c_F c_k^2 + c_S c_k + c_S c_F)\frac{g^2}{16m^3}|{\bf k}_1|^2
( (\bfsigma\times{\bf n}_1)^i {\bf n}_1^j - (\bfsigma\times{\bf n}_2)^j {\bf n}_2^i )[T^a,T^b]_{\alpha\beta} $$
$$-  (2c_{W_1} - 2c_{W_2} - c_S c_F + c_S c_k - 2c_F c_k^2)
\frac{g^2}{16m^3}|{\bf k}_1|^2 ( (\bfsigma\times{\bf n}_1)^i {\bf n}_1^j + (\bfsigma\times{\bf n}_2)^j {\bf n}_2^i)\{T^a,T^b\}_{\alpha\beta}$$
$$+  ( c_{B_2} + c_{B_1} -2c_{W_1} - c_S c_F -c_S c_k)\frac{g^2}{8m^3}|{\bf k}_1|^2 \bfsigma^k \epsilon^{ijk} [T^a,T^b]_{\alpha\beta}$$
$$ - c_{p'p}\frac{g^2}{16m^3}|{\bf k}_1|^2
[(  ({\bf n}_1\times{\bf n}_2)^j \bfsigma^i - ({\bf n}_1\times{\bf n}_2)^i \bfsigma^j
+ \epsilon^{ijk}({\bf n}_1-{\bf n}_2)^k(\bfsigma\cdot({\bf n}_1+{\bf n}_2)))\{T^a,T^b\}_{\alpha\beta} $$
$$ - (  ({\bf n}_1\times{\bf n}_2)^j \bfsigma^i + ({\bf n}_1\times{\bf n}_2)^i\bfsigma^j 
 - \epsilon^{ijk}({\bf n}_1 + {\bf n}_2)^k(\bfsigma\cdot({\bf n}_1 + {\bf n}_2)))[T^a,T^b]_{\alpha\beta}]$$
$$ + c_S c_k\frac{g^2}{8m^3}|{\bf k}_1|^2(1+ {\bf n}_1\cdot{\bf n}_2)\bfsigma^k \epsilon^{ijk} [T^a,T^b]_{\alpha\beta}$$
$$ +  c_F^2\frac{g^2}{8m^3} |{\bf k}_1|^2(1 + {\bf n}_1\cdot{\bf n}_2)
 ( (\bfsigma\cdot{\bf n}_2) {\bf n}_1^k \epsilon^{ijk}  + \bfsigma^j ({\bf n}_1\times{\bf n}_2)^i )[T^a,T^b]_{\alpha\beta}$$
$$- c_F c_k \frac{g^2}{4m^3}|{\bf k}_1|^2 (1 + {\bf n}_1\cdot{\bf n}_2) (\bfsigma\times{\bf n}_2)^j {\bf n}_2^i  [T^a,T^b]_{\alpha\beta}$$
$$ - \frac{g^2}{8m^3} |{\bf k}_1|^2(1 + {\bf n}_1\cdot{\bf n}_2)
[c_S((\bfsigma\times{\bf n}_2)^j {\bf n}_2^i + (\bfsigma\cdot{\bf n}_2){\bf n}_2^k \epsilon^{ijk})
+ c_F^2( (\bfsigma\cdot{\bf n}_2) {\bf n}_1^k \epsilon^{ijk}  + \bfsigma^j ({\bf n}_1\times{\bf n}_2)^i )$$
$$- 2c_F c_k(\bfsigma\times{\bf n}_2)^j {\bf n}_2^i] \{T^a,T^b\}_{\alpha\beta}$$
$$+ \frac{g^2}{8m^2}|{\bf k}_1|[(2c_F c_k - c_S c_k)((\bfsigma\times{\bf n}_1)^i {\bf n}_1^j - (\bfsigma\times{\bf n}_2)^j {\bf n}_2^i)
+ c_S c_k( (\bfsigma\cdot{\bf n}_1){\bf n}_1^k \epsilon^{ijk} +(\bfsigma\cdot{\bf n}_2){\bf n}_2^k \epsilon^{ijk})$$
$$+ c_F^2( (\bfsigma\cdot{\bf n}_1) {\bf n}_2^k \epsilon^{ijk}  + (\bfsigma\cdot{\bf n}_2) {\bf n}_1^k \epsilon^{ijk}
+ \bfsigma^i ({\bf n}_1\times{\bf n}_2)^j + \bfsigma^j ({\bf n}_1\times{\bf n}_2)^i)] \{T^a, T^b\}_{\alpha\beta}$$
$$+  c_F c_k \frac{g^2}{4m^2} |{\bf k}_1|((\bfsigma\times{\bf n}_1)^i {\bf n}_1^j + (\bfsigma\times{\bf n}_2)^j {\bf n}_2^i)[T^a,T^b]_{\alpha\beta}$$
$$+ c_F\frac{g^2}{2m}\bfsigma^k \epsilon^{ijk}[T^a,T^b]_{\alpha\beta}$$
$$+ c_F \frac{g^2}{4m} \frac{1}{1+{\bf n}_1\cdot{\bf n}_2}
(( \bfsigma\times{\bf n}_2)^i {\bf n}_1^j 
 - (\bfsigma\times{\bf n}_1)^j {\bf n}_2^i 
 + \bfsigma^i ({\bf n}_1\times{\bf n}_2)^j 
 + \bfsigma^j ({\bf n}_1\times{\bf n}_2)^i$$
\begin{equation}
 + 2(\bfsigma\times{\bf n}_1)^i {\bf n}_1^j 
 - 2(\bfsigma\times{\bf n}_2)^j {\bf n}_2^i)[T^a,T^b]_{\alpha\beta}\,.
\end{equation}
Note that $c_D$ does not appear explicitely. One can also observe that two combinations always appear in the observable: 
$\bar c_W \equiv c_{W_1}-c_{W_2}$ and $\bar c_{B_1}\equiv c_{B_1}-2c_{W_1}$. These, together with $c_{B_2}$ and $c_{p'p}$, are physical combinations,
i.e. they are gauge independent. This implies that the renormalization group equations (RGE) of these physical combinations can only depend 
on physical combinations of Wilson coefficients. Later on we will see that this is indeed the case. We suspect that individually 
$c_{W_1}$, $c_{W_2}$ and $c_{B_1}$ are gauge dependent quantities, since we are in agreement with Ref. \cite{Balzereit:1998jb}, where the calculation was done 
in Feynman gauge, at the level of single logs for physical combinations but we disagree for each of these three individually. 

For QED we obtain

$$\mathcal{A}^{ij} = \frac{g^2}{4m^2}|{\bf k}_1|[(2c_F c_k - c_S)((\bfsigma\times{\bf n}_1)^i {\bf n}_1^j - (\bfsigma\times{\bf n}_2)^j {\bf n}_2^i)
+ c_S( (\bfsigma\cdot{\bf n}_1){\bf n}_1^k \epsilon^{ijk} +(\bfsigma\cdot{\bf n}_2){\bf n}_2^k \epsilon^{ijk})$$
$$+ c_F^2( (\bfsigma\cdot{\bf n}_1) {\bf n}_2^k \epsilon^{ijk}  + (\bfsigma\cdot{\bf n}_2) {\bf n}_1^k \epsilon^{ijk}
+ \bfsigma^i ({\bf n}_1\times{\bf n}_2)^j + \bfsigma^j ({\bf n}_1\times{\bf n}_2)^i)]$$
$$-  (2c_{W_1} - 2c_{W_2} - c_S c_F + c_S c_k - 2c_F c_k^2)
\frac{g^2}{8m^3}|{\bf k}_1|^2 ( (\bfsigma\times{\bf n}_1)^i {\bf n}_1^j + (\bfsigma\times{\bf n}_2)^j {\bf n}_2^i)$$
$$ -c_{p'p}\frac{g^2}{8m^3}|{\bf k}_1|^2
(({\bf n}_1\times{\bf n}_2)^j \bfsigma^i - ({\bf n}_1\times{\bf n}_2)^i \bfsigma^j
+ \epsilon^{ijk}({\bf n}_1-{\bf n}_2)^k(\bfsigma\cdot({\bf n}_1+{\bf n}_2)))$$
$$ - \frac{g^2}{4m^3} |{\bf k}_1|^2(1 + {\bf n}_1\cdot{\bf n}_2)
[c_S(( \bfsigma\times{\bf n}_2)^j {\bf n}_2^i + (\bfsigma\cdot{\bf n}_2){\bf n}_2^k \epsilon^{ijk})
+ c_F^2( (\bfsigma\cdot{\bf n}_2) {\bf n}_1^k \epsilon^{ijk}  + \bfsigma^j ({\bf n}_1\times{\bf n}_2)^i )$$
\begin{equation}
 - 2c_F c_k(\bfsigma\times{\bf n}_2)^j {\bf n}_2^i]\,.
\end{equation}
Note that there is no ${\cal O}(1/m^0,1/m)$ contribution. Setting the Wilson coefficients to their tree level values we obtain

$$\mathcal{A}^{ij} = \frac{g^2}{4m^2}|{\bf k}_1|[ (\bfsigma\times{\bf n}_1)^i {\bf n}_1^j - (\bfsigma\times{\bf n}_2)^j {\bf n}_2^i
+  (\bfsigma\cdot{\bf n}_1){\bf n}_1^k \epsilon^{ijk} + (\bfsigma\cdot{\bf n}_2){\bf n}_2^k \epsilon^{ijk}$$
$$+  (\bfsigma\cdot{\bf n}_1) {\bf n}_2^k \epsilon^{ijk}  + (\bfsigma\cdot{\bf n}_2) {\bf n}_1^k \epsilon^{ijk}
+ \bfsigma^i ({\bf n}_1\times{\bf n}_2)^j + \bfsigma^j ({\bf n}_1\times{\bf n}_2)^i]$$
\begin{equation}
 + \frac{g^2}{4m^3} |{\bf k}_1|^2(1 + {\bf n}_1\cdot{\bf n}_2)
( - \bfsigma^j ({\bf n}_1\times{\bf n}_2)^i -  (\bfsigma\cdot{\bf n}_2) {\bf n}_1^k \epsilon^{ijk}  
+ (\bfsigma\times{\bf n}_2)^j {\bf n}_2^i - (\bfsigma\cdot{\bf n}_2){\bf n}_2^k \epsilon^{ijk})\,.
\end{equation}
This expression agrees with Eq. (19) in Ref. \cite{Balk:1993ev}.

The above analysis gives us the set of Wilson coefficients and their combinations that appear in physical observables: $\{\bar c_{W},\bar c_{B_1},c_{B_2},c_{p'p}\}$. 
We compute the anomalous dimensions for these, but also for the unphysical set: $\{c_{W_1}, c_{W_2},c_{B_1},c_{B_2},c_{p'p}\}$, in 
Coulomb gauge, as it can be important for future research investigating the possible gauge dependence of these Wilson coefficients.

\section{Anomalous dimensions for $1/m^3$ spin-dependent operators}
\label{Sec:RGE}

In this section we determine the anomalous dimensions of the Wilson coefficients associated to $1/m^3$ spin-dependent operators with ${\cal O}(\al)$ 
accuracy. In principle, one would like to only compute irreducible diagrams. However, as indicated in Ref. \cite{chromopolarizabilities}, this would involve 
considering a more extensive basis of operators, including those that vanish on shell. Instead, since we want to work in a minimal 
basis of operators, we will also need to consider reducible diagrams in a computation that resembles that of an S-matrix element. In particular, we will compute the divergent part of the amplitude for elastic scattering of the heavy quark with a 
tranverse gluon at one-loop. These divergences cancel with the divergences of the Wilson coefficients determining the anomalous dimension. The computation 
is organized in powers of $1/m$, up to ${\cal O}(1/m^3)$, by considering all possible insertions of the HQET Lagrangian operators. As a 
cross-check, we will also compute the elastic scattering of a heavy quark with a longitudinal gluon, which allows us to determine the anomalous dimension 
of the combination $c_{B_1}+c_{B_2}$. Furthermore, we compute the one transverse gluon-matrix element of the heavy quark, which allows us to cross-check the 
anomalous dimensions of 
$c_{W_1}$, $c_{W_2}$ and $c_{p'p}$. In the latter only irreducible diagrams enter the calculation. Note that $c_{B_1}$ has not been cross-checked in an 
independent calculation because one would need to consider irreducible and reducible diagrams with at least three external transverse gluons. Such a 
calculation is very hard and arguably not worth it because obtaining the correct structure of the $c_{B_1}$ vertex is non-trivial enough to be considered 
as a strong cross-check. In general, external gluons and heavy quarks will be considered to be on-shell i.e. free asymptotic states, so the free 
equations of motion (EOM) will be used throughout.

Note that we keep explicit the Wilson coefficients of the kinetic term for tracking purposes even though they are 
protected by reparametrization invariance ($c_k=c_4=1$ to any order in perturbation theory) \cite{Luke:1992cs}. Also, 
$c_{p'p}=c_F-1$ and the physical combination $\bar c_W=c_{W_1}-c_{W_2}=1$ are fixed by reparametrization invariance \cite{Manohar:1997qy}. We will 
check by explicit calculation that these relations are satisfied at LL.

The Coulomb gauge will be used throughout this paper. On the one hand, this significantly reduces the number of diagrams but, on the other hand,
the complexity of each one of them increases. It also makes it difficult to use standard routines for computations of diagrams designed 
for Feynman gauges and relativistic setups. However, since we are only looking for the pole, the calculation is feasible. The normalization of 
the heavy quark field, gluon fields, the strong coupling $g$ and the Wilson coefficients $c_F$ and $c_S$ are needed. In the Coulomb gauge, they 
read (we define $D=4+2\epsilon$):
\bea
\nn
Z_{A^0}^{-1/2}=Z_g&=&1+\frac{11}{6}C_A\frac{\al}{4\pi}\frac{1}{\epsilon} -\frac{2}{3}T_Fn_f\frac{\al}{4\pi}\frac{1}{\epsilon} \,,
\quad Z_{\bf A}^{\frac{1}{2}}=1-\frac{1}{2}C_A\frac{\al}{4\pi}\frac{1}{\epsilon}-\frac{2}{3}T_Fn_f\frac{\al}{4\pi}\frac{1}{\epsilon} 
\,,
\\
\nn
Z^2_gZ_{\bf A}&=&1+\frac{8}{3}C_A\frac{\al}{4\pi}\frac{1}{\epsilon}
\,,
\quad 
Z_l=1+C_F\frac{\al}{4\pi}\frac{1}{\epsilon}
\,, 
\quad
Z_h=1+\frac{{\bf p}^2}{m^2}\frac{4}{3}C_F\frac{\al}{4\pi}\frac{1}{\epsilon}
\,,
\eea
\begin{equation}
 c_{F,\,B}= c_{F,R} - c_{F,R} C_A\frac{\alpha}{4\pi}\frac{1}{\epsilon}\,,\;\;\;\;
 c_{S,\,B}= c_{S,R} - 2c_{F,R} C_A\frac{\alpha}{4\pi}\frac{1}{\epsilon}
\end{equation}
where
\begin{eqnarray}
  \cf=\frac{N_c^2-1}{2N_c}=\frac{4}{3}\,, \qquad C_A = N_c=3\,.
\end{eqnarray}
The subscript $B$ stands for bare and $R$ for renormalized quantities. Often the subscript $R$ will be removed in the following when it is 
understood. 

\subsection{QCD}
\label{sec:RGeq}

Let's consider the general case of QCD. First of all, for the pure gluonic sector, we have that $c_1^g$ is NLL, so it can be neglected.

\begin{figure}[!htb]  
	\includegraphics[width=0.95\textwidth]{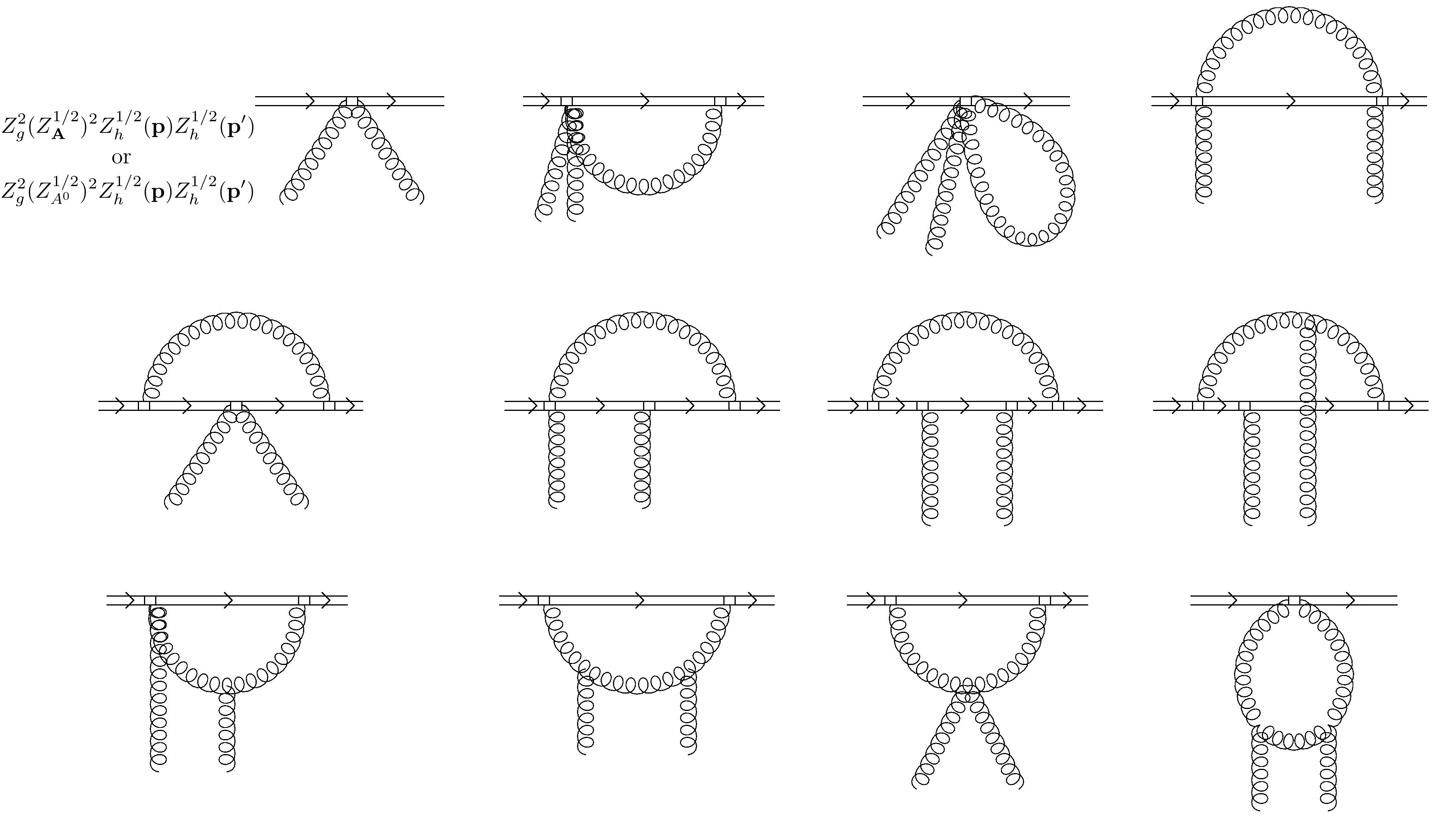}
	\includegraphics[width=0.95\textwidth]{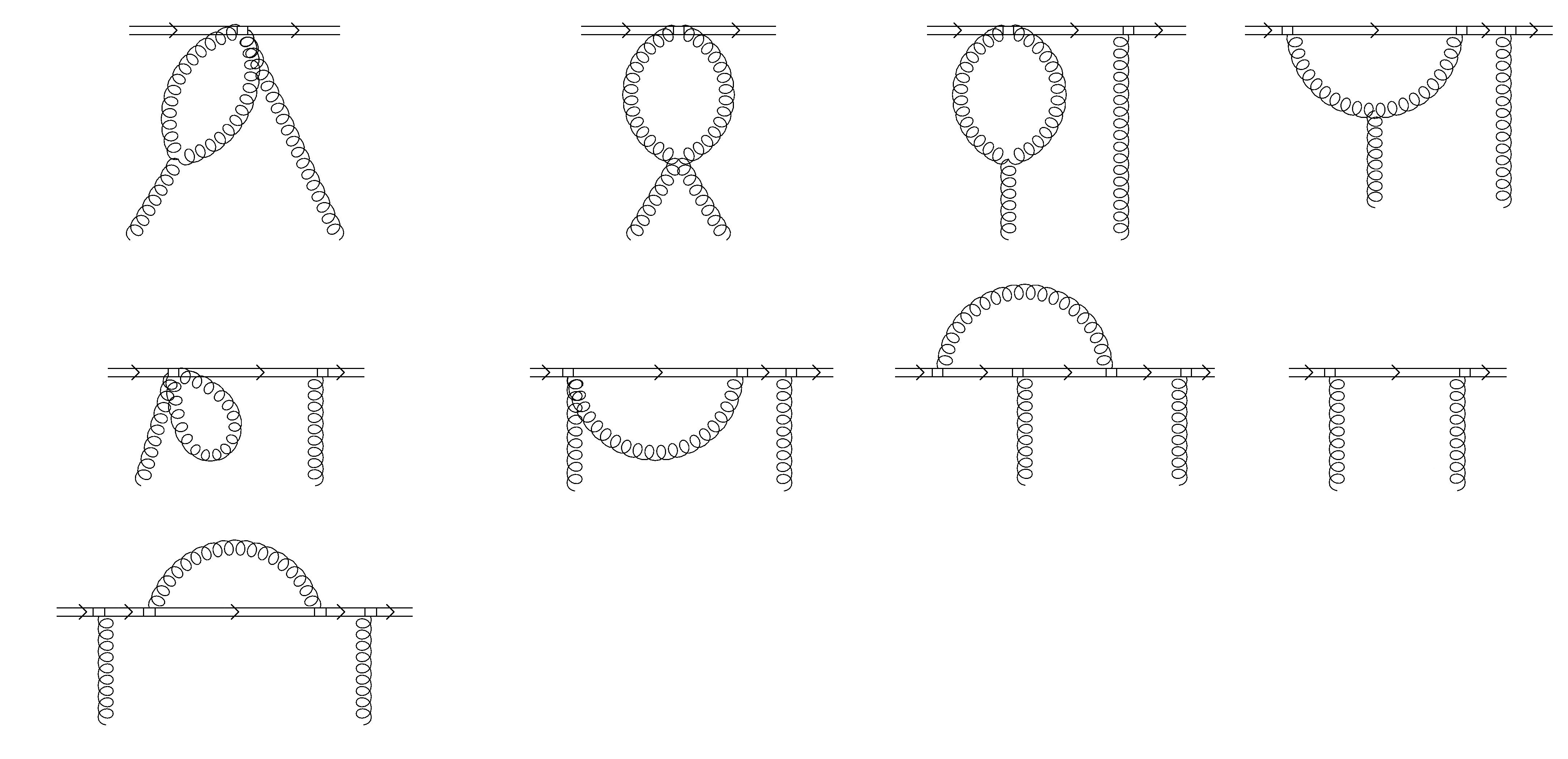}
\caption{Topologies contributing to the anomalous dimensions of the Wilson coefficients associated to $1/m^3$ spin-dependent operators in QCD. 
The double-line represent the heavy fermion, whereas the curly line represents either a transverse or a longitudinal 
gluon. Both external gluons are transverse or longitudinal depending on the kind of scattering we are considering. All diagrams are generated 
from these topologies by considering all possible vertices up to ${\cal O}(1/m^3)$. Tree level diagrams should be understood 
to be multiplied by Wilson coefficient, field and strong coupling counterterms.
\label{PlotsHeavygluon1}}   
\end{figure}

The running of the set: $\{c_{W_1},c_{W_2},c_{B_1},c_{B_2}, c_{p'p}\}$ is determined from the topologies drawn in Fig.~\ref{PlotsHeavygluon1}. From 
these, we generate all possible diagrams up to order $1/m^3$ by considering all possible vertices to the appropiate order in $1/m$ and/or kinetic 
insertions, which correspond to the expansion of the non-static heavy quark propagator. Note that diagrams of lower order than $1/m^3$ must also 
be considered, at least those that depend on the energy, as the use of the heavy quark EOM, $E= c_k \frac{{\bf p}^2}{2m}$, adds extra powers of $1/m$. This 
generates around 200 diagrams (without taking into account permutations and crossing) in both cases: the elastic scattering with a transverse gluon and, 
similarly, with a longitudinal gluon.

In the case of scattering with a transverse gluon, for diagrams proportional to $1/m^3$ operators, only the irreducible ones need to be considered. Note 
that this is not true for the case of scattering with a longitudinal gluon because the Coulomb vertex does not add extra powers of $1/m$. When one 
considers diagrams proportional to iterations of $1/m^2$ and/or $1/m$ operators one also has to consider reducible diagrams in both cases. One has 
to keep in mind that Taylor expanding reducible diagrams in the energy can produce non-local terms which cancel at the end of the calculation and 
all divergences can be absorbed by local counterterms that correspond to operators 
of the Lagrangian\footnote{If we only compute irreducible diagrams we would need a larger number of operators, in particular those that vanish on shell.}. 
It is also worth mentioning that we find that the sum of all reducible diagrams whose sub-irreducible diagram is $1/m$ or below 
($1/m^2$ or below in the case of the scattering with a longitudinal gluon) cancel with the renormalization of the tree level reducible diagrams. 
Therefore, non-local terms coming from expanding these diagrams in the energy vanish at all orders in the expansion.

\begin{figure}[!htb]    
	\includegraphics[width=0.95\textwidth]{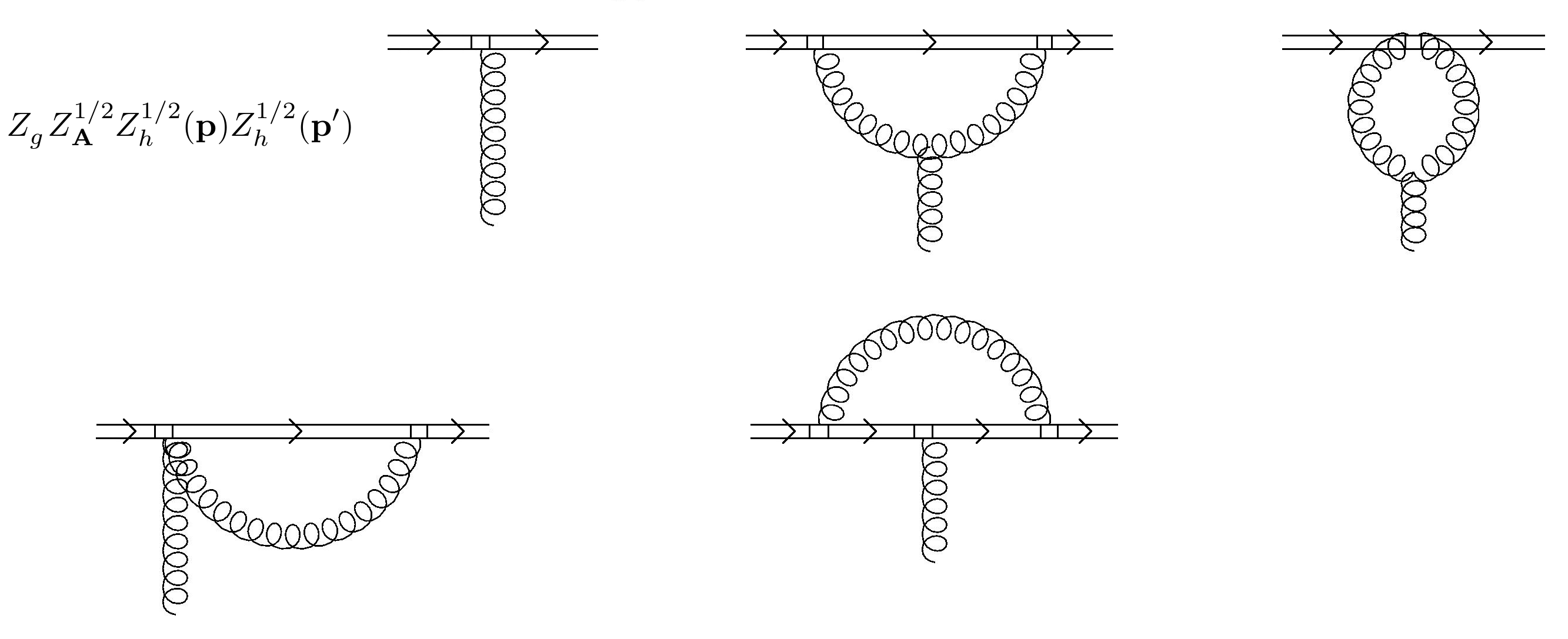}
\caption{Topologies contributing to the one tranverse gluon exchange. All diagrams are generated from these topologies by considering all possible 
vertices up to ${\cal O}(1/m^3)$. While the external gluon is transverse, internal gluons must be understood as either longitudinal 
or transverse.}
\label{plot1t}   
\end{figure}

Let's consider the calculation of the one tranverse gluon exchange, which has a peculiarity which deserves a comment. This calculation allows us to 
determine the anomalous dimensions of $c_{W_1}$, $c_{W_2}$, $c_{p'p}$ and $c_S$. The necesary topologies to produce 
the diagrams are shown in Fig. \ref{plot1t}. They produce around 50-60 diagrams without counting inverted ones. Note that we can only draw irreducible 
diagrams in this case.  What is 
interesting in this calculation is that one obtains a structure which does not look like any structure of the $1/m^3$ operators, i.e. the $1/m^3$ vertices 
with a single transverse gluon. So at first sight, 
it would look like a problem, since the divergence proportional to this structure could not be absorbed by any operator in the theory (leading one to suspect that there might be operators missing). However, this is not the case. The explanation is the following: in principle, one would consider 
$c_S$ as an $\mathcal{O}(1/m^2)$ operator. 
Nevertheless, the vertex with an external tranverse gluon is proportional to $k^0$, so it becomes $\mathcal{O}(1/m^3)$ after using the EOM. Therefore, in 
order to determine the running of $c_S$ through the calculation of the one gluon exchange one must consider this operator 
as an $\mathcal{O}(1/m^3)$ operator. Only in this way is the correct running of $c_S$ (expected from reparametrization invariance) obtained. So everything must 
be made physical, meaning put on shell, in order to arrive to proper results. Note that the runnig of $c_S$ will appear also 
in the determination of the running of Wilson coefficients at higher orders in $1/m$ if it is done through the calculation of the 
one tranverse gluon matrix element of a heavy quark, because the EOM have corrections in $1/m$. In particular, it will appear at $\mathcal{O}(1/m^5)$. 
This is important to keep in mind in future calculations.

The renormalization group equations for the unphysical set $\{c_{W_1},c_{W_2},c_{B_1},c_{B_2}, c_{p'p}\}$ in Coulomb gauge read:

$$\nu\frac{d}{d\nu}c_{W_1}=  \frac{\alpha}{\pi}\bigg(\frac{13}{12}c_{W_1}C_A  + \frac{7}{12}c_{W_2}C_A 
- \frac{1}{4} c_{B_1} C_A - \frac{1}{8} c_{B_2} C_A 
+ \frac{1}{24}c_{p'p} C_A $$
\begin{equation}
 + \frac{7}{24}c_S c_k C_A  - \frac{1}{6}c_S c_F C_A 
 - \frac{1}{12}c_F c_k^2 (16C_F + 15C_A) + \frac{7}{8}c_F^2 c_k C_A \bigg)\,,
\end{equation}

$$\nu\frac{d}{d\nu}c_{W_2}= \frac{\alpha}{\pi} \bigg(\frac{7}{12}c_{W_1}C_A  + \frac{13}{12}c_{W_2}C_A 
- \frac{1}{4} c_{B_1} C_A  - \frac{1}{8} c_{B_2} C_A 
+ \frac{1}{24}c_{p'p} C_A $$
\begin{equation}
 - \frac{5}{24}c_S c_k C_A  - \frac{1}{6}c_S c_F C_A 
 - \frac{1}{12}c_F c_k^2 (16C_F + 3C_A) + \frac{7}{8}c_F^2 c_k C_A \bigg)\,,
\end{equation}

$$\nu\frac{d}{d\nu} c_{B_1}=   \frac{\alpha}{\pi}\bigg( \frac{1}{6} c_{W_1} C_A + \frac{1}{6} c_{W_2} C_A 
+ c_{B_1}C_A -  \frac{1}{3}c_{B_2} C_A  
+ \frac{7}{12} c_{p'p} C_A$$
\begin{equation}
 + \frac{1}{12} c_S c_k C_A - \frac{1}{4} c_S c_F C_A 
 + \frac{7}{6} c_F c_k^2 C_A  + \frac{7}{6} c_F^2 c_k C_A + \frac{3}{2} c_F^3 C_A \bigg)\,,
\end{equation}

$$\nu\frac{d}{d\nu} c_{B_2}
=   \frac{\alpha}{\pi} \bigg( c_{W_2} C_A 
- \frac{1}{2}c_{B_1} C_A 
+ \frac{7}{6}c_{B_2} C_A$$
\begin{equation}
 - \frac{4}{3} c_S c_k (4C_F + C_A) 
 - \frac{1}{6} c_S c_F C_A
 + \frac{4}{3} c_F c_k^2 (2C_F - C_A)
 + \frac{2}{3} c_F^2 c_k C_A
 - \frac{3}{2} c_F^3 C_A \bigg)\,,
\end{equation}

\begin{equation}
 \nu\frac{d}{d\nu}c_{p'p}= \frac{\alpha}{\pi}\bigg(\frac{1}{2}c_{p'p} C_A  - \frac{1}{2}c_S c_k C_A + c_F c_k^2 C_A \bigg)\,.
\end{equation}
The renormalization group equations for the physical set $\{\bar c_{W}, \bar c_{B_1}, c_{B_2}, c_{p'p}\}$ read:

\begin{equation}
\label{cWbarrge}
 \nu\frac{d}{d\nu}\bar c_{W} = \frac{\alpha}{\pi} \bigg( \frac{1}{2}\bar c_{W} C_A  
+ \frac{1}{2}c_S c_k C_A - c_F c_k^2  C_A \bigg)=0\,,
\end{equation}

$$\nu\frac{d}{d\nu} \bar c_{B_1} = 
 \frac{\alpha}{\pi} \bigg( \frac{3}{2} \bar c_{B_1} C_A 
+ \bar c_{W}C_A 
- \frac{1}{12}c_{B_2} C_A  
+ \frac{1}{2} c_{p'p} C_A$$
\begin{equation}
 - \frac{1}{2}c_S c_k C_A 
 + \frac{1}{12} c_S c_F C_A 
 + \frac{1}{3}c_F c_k^2 (8C_F + 11C_A)
 - \frac{7}{12} c_F^2 c_k C_A 
 + \frac{3}{2} c_F^3 C_A \bigg)\,,
\end{equation}

$$\nu\frac{d}{d\nu} c_{B_2}
 = \frac{\alpha}{\pi}  \bigg( - \frac{1}{2}\bar c_{B_1} C_A 
- \bar c_W C_A 
+ \frac{7}{6}c_{B_2} C_A $$
\begin{equation}
 - \frac{4}{3} c_S c_k (4C_F + C_A)
 - \frac{1}{6} c_S c_F C_A
 + \frac{4}{3} c_F c_k^2 (2C_F - C_A) 
 + \frac{2}{3} c_F^2 c_k C_A
 - \frac{3}{2} c_F^3 C_A \bigg)\,,
\end{equation}

\begin{equation}
 \nu\frac{d}{d\nu}c_{p'p}= \frac{\alpha}{\pi}\bigg(\frac{1}{2}c_{p'p} C_A  - \frac{1}{2}c_S c_k C_A + c_F c_k^2 C_A \bigg)\,.
\end{equation}
Where $\bar c_{W}$ and $\bar c_{B_1}$ come from the definitions given in Sec. \ref{Sec:Compton}. The last equality in Eq. (\ref{cWbarrge}) can be easily 
deduced by using the relations between Wilson coefficients imposed by reparametrization invariance. When writing the counterterm of each Wilson coefficient 
it is enough to know that the scaling with the renormalization scale is $\nu^{2\epsilon}$. It is quite remarkable that the RG equations depend only on 
gauge-independent combinations of Wilson coefficients: $\bar c_{W}$, $\bar c_{B_1}$ and $c_{B_2}$. This is quite a strong check, as at intermediate 
steps we get contributions from $c_{W_1}$, $c_{W_2}$ and $c_{B_1}$, which only at the end of the computation arrange themselves in gauge-independent combinations.

\clearpage
\subsection{QED}
\label{sec:RGeqqed}

In this section we analyze the purely abelian case of QED. To do this, we just need to take the appropriate limit
of the results found in Sec. \ref{sec:RGeq} i.e. to take $C_F=1$, $C_A=0$ and $n_f=0$. Note that the operators proportional to $c_{B_1}$ and 
$c_{B_2}$ do not appear now. The diagrams that contribute are the same drawn in Fig.~\ref{PlotsHeavygluon1} and Fig.~\ref{plot1t} disregarding non-abelian 
diagrams.

The renormalization group equations read:

\begin{equation}
 \nu\frac{d}{d\nu}c_{W_1}=  - \frac{4}{3}c_F c_k^2 C_F \frac{\alpha}{\pi} \,,\;\;\;\;\;
 \nu\frac{d}{d\nu}c_{W_2}= - \frac{4}{3}c_F c_k^2 C_F \frac{\alpha}{\pi}\,,\;\;\;\;\;
 \nu\frac{d}{d\nu}c_{p'p}=0\,.
\end{equation}
This result is in agreement with the explicit single logs given in Refs. \cite{Balzereit:1998jb,Balzereit:1998vh}, where the calculation was 
done in Feynman gauge. This is not that
strange; for instance,  the running of $c_D$ in QED happens to be equal in the Coulomb and Feynman gauges (see the discussion in 
Ref. \cite{Pineda:2001ra}). The analysis done in Sec. \ref{Sec:Compton} suggests that the physical object is still $\bar c_W$, though. The results 
obtained are in agreement with reparametrization invariance relations given in Ref. \cite{Manohar:1997qy} (recall that $c_F$ has no running in QED).

\section{Solution and numerical analysis}
\label{Sec:numerics}

The RG equations obtained in the previous section depend on a list of five Wilson coefficients 
${\bf A}=\{ c_{W_1}, c_{W_2}, c_{B_1}, c_{B_2},c_{p'p} \}$. The running of $c_{p'p}$ has been found to be the same as that of $c_F$, as predicted 
by reparameterization invariance, so we will not give an explicit expression. The remaining equations can be written more compactly
in a matrix form 

\be
\nu \frac{d}{d \nu} {\bf A}= \frac{\al}{\pi}({\bf M}  {\bf A}+{\bf F}(\al))
\,.
\ee
The matrix ${\bf M}$ follows from the results of the previous section. We only need the running of $\al$ with LL accuracy:
\be
 \nu   {{\rm d} \over {\rm d} \nu}   \al
\equiv
\beta(\alpha_s)
=
-2 \al
\left\{\beta_0{\al \over 4 \pi} + \cdots\right\}
\,,
\ee
where 
\begin{eqnarray}
  \beta_0&=&{11 \over 3}C_A -{4 \over 3}T_Fn_f\,,
\end{eqnarray}
and $n_f$ is the number of dynamical (active) quarks i.e. the number of light quarks. Recall that $n_f=0$ for the case we are studying.
 
In this approximation, the above equation can be simplified to
\be
\frac{d}{d \al} {\bf A}= -\frac{2}{ \beta_0\al}({\bf M}  {\bf A} + {\bf F}(\al))
\,.
\ee
It is useful to define $z \equiv \left(\frac{\al(\nu)}{\al(m)}\right)^{\frac{1}{\beta_0}}\simeq
1-\frac{1}{2\pi}\al(\nu)\ln (\frac{\nu}{m})$ and write the equation above as:

\begin{equation}
 \frac{d{\bf A}}{dz}=-\frac{2}{z}(\mathbf{M}{\bf A} + {\bf F}(z))\,.
\end{equation}
We also need the initial matching conditions at the hard scale, at tree-level. They have been determined 
in Ref.~\cite{Manohar:1997qy} and read $c_k=c_F=c_S=c_{W_1}=c_{B_1}=1$
and $c_{W_2}=c_{p'p}=c_{B_2}=0$.

Note that the matching coefficient of the
kinetic term is protected by reparametrization invariance
($c_k=1$ to any order in perturbation theory) \cite{Luke:1992cs}.
Nevertheless, although we set them to 1 when solving the RG equations, we keep them explicit in the RG equations for
tracking purposes. 

After solving the RG  equations we obtain the LL running of the Wilson coefficients associated to the $1/m^3$ spin-dependent operators 
of the HQET Lagrangian. We obtain the following analytic results 
for the unphysical set in Coulomb gauge:

$$ c_{W_1} = \frac{47}{82} - \frac{160C_F}{287C_A}
+ \left(\frac{1}{3}  + \frac{10C_F}{3C_A}\right)z^{-C_A}
- z^{-2C_A}
+ \frac{9}{2}z^{-3C_A}
+ \left(5 + \frac{22C_F}{7C_A}\right)z^{-\frac{7}{3}C_A}$$
$$+ \left(- \frac{517}{123}
 + \frac{14}{41}\sqrt{\frac{2}{5}}
 - \sqrt{\frac{5}{2}} 
 + \frac{50}{123}\sqrt{10}
 - \frac{364C_F}{123C_A}
 - \frac{452C_F}{123C_A}\sqrt{\frac{2}{5}}\right)z^{\frac{1}{6}(-16+\sqrt{10})C_A}$$
 \begin{equation}
 + \left( - \frac{517}{123} 
 -\frac{14}{41}\sqrt{\frac{2}{5}} 
 + \sqrt{\frac{5}{2}}
 -\frac{50}{123}\sqrt{10}
 - \frac{364C_F}{123C_A}
 + \frac{452C_F}{123C_A}\sqrt{\frac{2}{5}}\right)z^{-\frac{1}{6}(16+\sqrt{10})C_A}\,,
\end{equation}\\

$$ c_{W_2} = -\frac{35}{82} - \frac{160C_F}{287C_A}
+ \left(\frac{1}{3}  + \frac{10C_F}{3C_A}\right)z^{-C_A}
- z^{-2C_A}
+ \frac{9}{2}z^{-3C_A}
+ \left(5 + \frac{22C_F}{7C_A}\right)z^{-\frac{7}{3}C_A}$$
$$+ \left(- \frac{517}{123}
 + \frac{14}{41}\sqrt{\frac{2}{5}}
 - \sqrt{\frac{5}{2}} 
 + \frac{50}{123}\sqrt{10}
 - \frac{364C_F}{123C_A}
 - \frac{452C_F}{123C_A}\sqrt{\frac{2}{5}}\right)z^{\frac{1}{6}(-16+\sqrt{10})C_A}$$
 \begin{equation}
 + \left( - \frac{517}{123} 
 -\frac{14}{41}\sqrt{\frac{2}{5}} 
 + \sqrt{\frac{5}{2}}
 -\frac{50}{123}\sqrt{10}
 - \frac{364C_F}{123C_A}
 + \frac{452C_F}{123C_A}\sqrt{\frac{2}{5}}\right)z^{-\frac{1}{6}(16+\sqrt{10})C_A}\,,
\end{equation}\\

$$ c_{B_1} = \frac{55}{123} - \frac{1184C_F}{861C_A} 
+ \left(- \frac{19}{9}  + \frac{44C_F}{9C_A}\right)z^{-C_A}
- z^{-2C_A} 
- 6z^{-3C_A} 
+ \left(10 + \frac{44C_F}{7C_A}\right)z^{-\frac{7}{3}C_A}$$
$$+ \left(- \frac{62}{369}
 - \frac{1034}{369}\sqrt{10}
 - \frac{1808C_F}{369C_A}
 - \frac{728C_F}{369C_A}\sqrt{10}\right)z^{\frac{1}{6}(-16+\sqrt{10})C_A}$$
 \begin{equation}
 + \left(- \frac{62}{369}
 + \frac{1034}{369}\sqrt{10}
 - \frac{1808C_F}{369C_A}
 + \frac{728C_F}{369C_A}\sqrt{10}\right)z^{-\frac{1}{6}(16+\sqrt{10})C_A}\,,
\end{equation}\\

$$c_{B_2} = -\frac{24}{41} - \frac{192C_F}{41C_A} 
+ \left(\frac{11}{3} + \frac{32C_F}{3C_A}\right)z^{-C_A}
+ z^{-2C_A} + 18z^{-3C_A} $$
$$+\left( - \frac{1358}{123}
 + \frac{64}{41}\sqrt{\frac{2}{5}}
 + \frac{298}{123}\sqrt{10} 
 - \frac{368C_F}{123C_A}
 + \frac{8C_F}{123C_A}\sqrt{\frac{2}{5}}\right)z^{\frac{1}{6}(-16+\sqrt{10})C_A} $$
\begin{equation}
 + \left( - \frac{1358}{123}
 - \frac{64}{41}\sqrt{\frac{2}{5}}
 - \frac{298}{123}\sqrt{10}
 - \frac{368C_F}{123C_A}
 - \frac{8C_F}{123C_A}\sqrt{\frac{2}{5}} \right) z^{-\frac{1}{6}(16+\sqrt{10})C_A}\,.
\end{equation}
The solution for the physical set of Wilson coefficients read:

\begin{equation}
\label{cWbar}
 \bar c_{W} = 1\,,
\end{equation}

$$\bar c_{B_1} = -\frac{86}{123} - \frac{32C_F}{123C_A}
- \left( \frac{25}{9} + \frac{16C_F}{9C_A} \right)z^{-C_A}
+ z^{-2C_A}
- 15z^{-3C_A} $$
$$+ \left( \frac{3040}{369} 
 - \frac{5077}{369}\sqrt{\frac{2}{5}}
 + \frac{376C_F}{369C_A}
 - \frac{928C_F}{369C_A}\sqrt{\frac{2}{5}}\right)z^{\frac{1}{6}(-16+\sqrt{10})C_A} $$
\begin{equation}
 + \left(\frac{3040}{369}
 + \frac{5077}{369}\sqrt{\frac{2}{5}}
 + \frac{376C_F}{369C_A} 
 + \frac{928C_F}{369C_A}\sqrt{\frac{2}{5}}\right)z^{-\frac{1}{6}(16+\sqrt{10})C_A} \,,
\end{equation}

$$c_{B_2} = -\frac{24}{41} - \frac{192C_F}{41C_A} 
+ \left(\frac{11}{3} + \frac{32C_F}{3C_A}\right)z^{-C_A}
+ z^{-2C_A} + 18z^{-3C_A} $$
$$+\left( - \frac{1358}{123}
 + \frac{64}{41}\sqrt{\frac{2}{5}}
 + \frac{298}{123}\sqrt{10} 
 - \frac{368C_F}{123C_A}
 + \frac{8C_F}{123C_A}\sqrt{\frac{2}{5}}\right)z^{\frac{1}{6}(-16+\sqrt{10})C_A} $$
\begin{equation}
 + \left( - \frac{1358}{123}
 - \frac{64}{41}\sqrt{\frac{2}{5}}
 - \frac{298}{123}\sqrt{10}
 - \frac{368C_F}{123C_A}
 - \frac{8C_F}{123C_A}\sqrt{\frac{2}{5}} \right) z^{-\frac{1}{6}(16+\sqrt{10})C_A}\,.
\end{equation}
As can be seen, Eq. (\ref{cWbar}) satifies reparametrization invariance. We have proven by explicit calculation that, at LL, $c_{p'p}$ and 
$\bar c_W$ satisfy the relations imposed by reparametrization invariance given in Ref. \cite{Manohar:1997qy}.

If we expand the above solutions in powers of $\al$ we can explicitely write the single log (it can also be obtained by trivial inspection of the
RG equations in Sec. \ref{Sec:RGE}). We obtain for the unphysical set that

\begin{equation}
\label{sl4}
c_{W_1} = 1 + \left( - \frac{4}{3}C_F + \frac{7}{12}C_A \right)\frac{\alpha}{\pi}\ln\left(\frac{\nu}{m}\right)+{\cal O}(\al^2)\,,
\end{equation}

\begin{equation}
\label{sl5}
c_{W_2} =  \left(  -  \frac{4}{3}C_F + \frac{7}{12}C_A \right)\frac{\alpha}{\pi}\ln\left(\frac{\nu}{m}\right)+{\cal O}(\al^2)\,,
\end{equation}

\begin{equation}
\label{sl6}
c_{B_1}= 1 + \frac{29}{6} C_A \frac{\alpha}{\pi}\ln\left(\frac{\nu}{m}\right)+{\cal O}(\al^2)\,,
\end{equation}

\begin{equation}
\label{sl7}
c_{B_2} = -\left(\frac{8}{3}C_F + \frac{25}{6} C_A\right) \frac{\alpha}{\pi}\ln\left(\frac{\nu}{m}\right)+{\cal O}(\al^2)\,,
\end{equation}
and for the physical set, that

\begin{equation}
\label{sl1}
 \bar c_{W} = 1 +{\cal O}(\al^2)\,,
\end{equation}

\begin{equation}
\label{sl2}
 \bar c_{B_1} = -1 + \left(  \frac{8}{3}C_F + \frac{11}{3} C_A \right)\frac{\alpha}{\pi}\ln\left(\frac{\nu}{m}\right)+{\cal O}(\al^2)\,,
\end{equation}

\begin{equation}
\label{sl3}
 c_{B_2} = -\left(\frac{8}{3}C_F + \frac{25}{6} C_A\right) \frac{\alpha}{\pi}\ln\left(\frac{\nu}{m}\right)+{\cal O}(\al^2)\,.
\end{equation}

\begin{figure}[!htb]
	%\begin{center}      
	\includegraphics[width=0.43\textwidth]{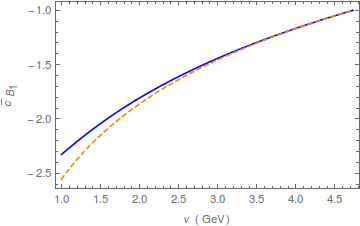}
	%\put(-45,-10){$c_{A_1}$} 
\hspace{6ex}
	\includegraphics[width=0.43\textwidth]{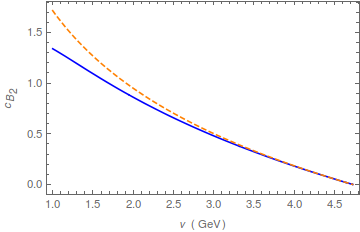}
	%\put(-45,-10){$\bar c_{A_2}$} 
	%
	\vspace{1ex}
	\includegraphics[width=0.43\textwidth]{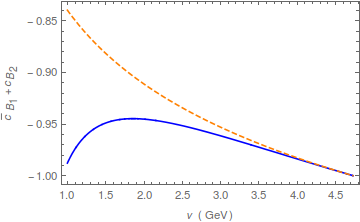}
\caption{Running of the physical $1/m^3$ spin-dependent Wilson coefficients. The continuous line is the LL result with $n_f=0$ and the dashed line is 
the single LL result which does not depend on $n_f$. 
\label{Plots}}   
%\end{center}
\end{figure}

In Fig. \ref{Plots} one can see the above results when applied to the bottom heavy quark case, ilustrating the importance of incorporating large 
logarithms in heavy quark physics. Only physical combinations and specific combinations that appear in physical observables, like Compton scattering, are 
represented. We run the Wilson coefficients from the heavy quark mass to 1 GeV. For illustrative purposes, we take $m_b=4.73$ GeV and $\al(m_b)=0.215943$.

Concerning the numerical analysis, we observe that the effect due to the logarithms is large in general (not for QED however, where the only physical 
combination that appears, $\bar c_W$, does not run). This is because the coefficients multiplying the 
logs are large, in particular those that multiply the non-abelian coefficient $C_A$. We also observe that the LL resummation is saturated 
by the single log in all cases except in the combination $\bar c_{B_1} + c_{B_2}$. Let us now discuss in more detail each individual Wilson coefficient. We 
observe the following: $\bar c_{B_1}$ changes from -1 to -2.3 after running. The case of $c_{B_2}$ is rather similar, it goes from 0 to 1.45 after running. In 
general, the effect of the resummation of logarithms is not quite large, but certainly sizable. It introduces a change of approximately 0.3 with 
respect to the single log result. For the combination $\bar c_{B_1} + c_{B_2}$ the effect is very small, it goes from -1 to -0.99 even though it has a maximum in 
which the value is -0.95. In this case the resummation of logs is important because the behaviour is not saturated by the single log.

\section{Comparison with earlier work}
\label{Sec:Comparison}

The LL running of the Wilson coefficients associated to the $1/m^3$ operators of the HQET Lagrangian without considering light fermion effects was first 
addressed in Refs. \cite{Balzereit:1998jb,Balzereit:1998vh}, where expressions for the anomalous dimension matrix and explicit 
expressions for the Wilson coefficients with single log accuracy are given. As in Ref. \cite{chromopolarizabilities} for the spin-independent 
case, we find that these results are mutually inconsistent, as the anomalous 
dimension matrix produces different expressions for the single log result compared to the explicit single log expressions written in these references.

The basis of operators these results were obtained from is different from the basis used in this paper, so in order to 
compare our results, we have to change the operator basis. This is done via field redefinitions, which at the order we are working in, is 
equivalent to using the full equations of motion to order $1/m$. To this purpose, we use the HQET Lagrangian in a general frame, 
Eq. (8) in Ref. \cite{Manohar:1997qy}\footnote{We take this opportunity to correct a missprint 
in the term proportional to $c_{p'p}$, where the minus sign appearing there should be a plus sign in order to reproduce the Lagrangian Eq. (6) and 
Eq. (7) in Ref. \cite{Manohar:1997qy}.}. We obtain the 
following relations between the spin-dependent $1/m^3$ Wilson coefficients in the two bases:

\begin{equation}
 c_5^{(3)}= - c_{B_2} - c_k c_F^2 - c_S c_F\,,
\end{equation}
\begin{equation}
 c_6^{(3)} =  -c_{W_1} - c_{p'p} + c_k^2 c_F + \frac{1}{2}c_D c_F + \frac{1}{2}c_S c_k\,,
\end{equation}
\begin{equation}
 c_7^{(3)}= 2c_{W_2} - 2c_{p'p} - c_{B_1} + c_k c_F^2 + c_S c_F \,,
\end{equation}
\begin{equation}
 c_8^{(3)} =  -c_{W_1} - c_{p'p} + c_k^2 c_F + \frac{1}{2}c_D c_F + \frac{1}{2}c_S c_k\,,
\end{equation}
\begin{equation}
 c_9^{(3)} =  -c_{B_1} + c_k c_F^2 + c_S c_F\,,
\end{equation}
\begin{equation}
 c_{10}^{(3)} =  c_{p'p} + c_{B_1} - c_k c_F^2 - c_S c_F\,,
\end{equation}
\begin{equation}
 c_{11}^{(3)} =  c_{p'p} + c_{B_1} - c_k c_F^2 - c_S c_F\,.
\end{equation}
Note that $c_6^{(3)} = c_8^{(3)}$ and $c_{10}^{(3)}=c_{11}^{(3)}$. This is to be expected since it is well-known from Ref. \cite{Manohar:1997qy} that 
there are five spin-dependent operators and five different Wilson coefficients, whereas in Refs. \cite{Balzereit:1998jb,Balzereit:1998vh} 
there are seven operatos and seven different Wilson coefficients. Thereby, one must find that two Wilson coefficients have to be equal to another two. 
In Ref. \cite{chromopolarizabilities} these relations between the Wilson coefficients  at $\mathcal{O}(1/m)$ and $\mathcal{O}(1/m^2)$ were also found:
\bea
 c_1^{(1)} = c_k
\,,\qquad
 c_2^{(1)} = c_F
\,, \qquad
 c_1^{(2)} = -c_D
\,, \qquad
 c_2^{(2)} = c_S
\,.
\eea
In Ref. \cite{chromopolarizabilities} expressions for the correct anomalous dimension matrix for the spin-independent Wilson coefficients in the basis 
used in Refs. \cite{Balzereit:1998jb,Balzereit:1998vh} were given. However the situation for the spin-dependent case turns out to be more complicated, since 
Wilson coefficients are gauge dependent. Because the calculation in these references was done in Feynman gauge and in this paper in Coulomb gauge, we can not 
give a prediction for the anomalous dimension matrix in Refs. \cite{Balzereit:1998jb,Balzereit:1998vh} in Feynman gauge (note that in the 
spin-independent case there were also gauge dependent Wilson coefficients, but all gauge dependence came from $c_D$ and $c_M$, for which expressions are 
well-known in Feynman gauge). Instead, to compare results, we compute the RG equations in our basis for physical quantities from the anomalous 
dimension matrices given in Ref. \cite{Balzereit:1998jb}. To do this, one needs the inverse relations between the Wilson coefficients in 
the two bases:

\begin{equation}
 c_{p'p} = c_9^{(3)} + c_{10}^{(3)}\,,
\end{equation}
\begin{equation}
 c_{W_1} = -c_6^{(3)} - c_9^{(3)} - c_{10}^{(3)} + c_1^{(1)\,2}c_2^{(1)} - \frac{1}{2}c_1^{(2)}c_2^{(1)} + \frac{1}{2}c_2^{(2)}c_1^{(1)}\,,
\end{equation}
\begin{equation}
 c_{W_2} = \frac{1}{2} c_7^{(3)} + \frac{1}{2}c_9^{(3)} + c_{10}^{(3)}\,,
\end{equation}
\begin{equation}
 c_{B_1} = -c_9^{(3)} + c_1^{(1)}c_2^{(1)\,2} + c_2^{(2)}c_2^{(1)}\,,
\end{equation}
\begin{equation}
 c_{B_2} = -c_5^{(3)} - c_1^{(1)}c_2^{(1)\,2} - c_2^{(2)}c_2^{(1)}\,.
\end{equation}
Firstly, note that the anomalous dimension matrix in Ref. \cite{Balzereit:1998jb} gives $c_6^{(3)}\neq c_8^{(3)}$. This already
disagrees with our results and with the explicit single log results given in Table II of that reference. We continue with the 
comparison nonetheless. We take the expression for $c_6^{(3)}$, which is the one which minimizes the discrepancies. The RG equations read

\begin{equation}
\label{cpp}
 \nu\frac{d}{d\nu} c_{p'p} =\frac{\alpha}{\pi}\left( C_A c_F c_k^2 + \frac{1}{2}C_A c_{p'p} - \frac{1}{2}C_A c_S c_k\right)\,,
\end{equation}

\begin{equation}
\label{cWrg}
 \nu\frac{d}{d\nu}\bar c_W = \frac{\alpha}{\pi}\bigg( {\bf\frac{1}{6}}C_A c_D c_F - {\bf\frac{5}{12}}C_A c_F^3 - {\bf\frac{3}{4}}C_A c_F^2 c_k 
 -{\bf\frac{1}{12}}C_A c_F c_k^2 - {\bf\frac{4}{3}}C_F c_F c_k^2 - {\bf\frac{1}{4}}C_A c_S c_k + \frac{1}{2}C_A \bar c_W\bigg)\,,
\end{equation}

$$ \nu\frac{d}{d\nu}\bar c_{B_1} = \frac{\alpha}{\pi}\bigg( \frac{3}{2}C_A \bar c_{B_1} - \frac{1}{12}C_A c_{B_2} -{\bf\frac{1}{3}}C_A c_D c_F 
 + {\bf\frac{5}{6}}C_A c_F^3 + {\bf\frac{35}{12}}C_A c_F^2 c_k + {\bf\frac{11}{6}}C_A c_F c_k^2 $$
\begin{equation}
\label{cb1rg}
  + {\bf\frac{16}{3}}C_F c_F c_k^2 + \frac{1}{2}C_Ac_{p'p} - {\bf\frac{17}{12}}C_A c_F c_S + {\bf 1}C_A c_S c_k + C_A \bar c_W\bigg)\,,
\end{equation}

$$\nu\frac{d}{d\nu}c_{B_2} = \frac{\alpha}{\pi}\bigg( -\frac{1}{2}C_A \bar c_{B_1} + \frac{7}{6}C_A c_{B_2} 
 - {\bf\frac{4}{3}}C_A c_F^2 c_k - \frac{4}{3}C_A c_F c_k^2 + \frac{8}{3}C_F c_F c_k^2 + {\bf\frac{4}{3}}C_A c_F c_S$$
\begin{equation}
  - \frac{4}{3}C_A c_S c_k -\frac{16}{3}C_F c_S c_k - C_A\bar c_{W} + {\bf 0} c_F^3\bigg)\,,
\end{equation}
where numbers in bold indicate a discrepancy with respect to our results. In general we find disagreement for all RG equations (even in QED) except  
for Eq. (\ref{cpp}), which satisfies reparametrization invariance. Conceptually, the diagreement with Eqs. (\ref{cWrg}-\ref{cb1rg}) is important, as 
these equations do not depend only on physical combinations of Wilson coefficients due to the explicit appearence of $c_D$, which is gauge dependent. 
In addition, Eq. (\ref{cWrg}) does not satisfy reparametrization invariance.

On the other hand, it is remarkable that using the single log results given in Table II of Ref. \cite{Balzereit:1998vh} one finds agreement with our single 
log results for the physical quantities $\bar c_W$, $\bar c_{B_1}$, $c_{B_2}$, given in Eqs. (\ref{sl1}-\ref{sl3}), and $c_{p'p}$, not presented 
explicitely. However, we find disagreement for the unphysical quantities $c_{W_1}$, $c_{W_2}$ and $c_{B_1}$, given in Eqs. (\ref{sl4}-\ref{sl6}). If we 
trust the explicit single logs presented in this reference, this is a clear indication these Wilson coefficients are gauge dependent.

\section{Conclusions}
\label{sec:conclusions}

We have computed the LL running of the Wilson coefficients associated to the spin-dependent $1/m^3$ operators of the HQET Lagrangian without light fermion 
effects. We observe that reparametrization invariance relations are satisfied and that the running of physical quantities depend only on gauge-independent 
quantities, as expected. Numerically, we observe that the running produces a large effect, except for the combination $\bar c_{B_1}+c_{B_2}$, which also 
appears in Compton scattering. However, in this case, the resummation of large logarithms also happens to be important because the behaviour of this combination 
is not saturated by the single log.

We have compared our results with the previous work done in Refs. \cite{Balzereit:1998jb,Balzereit:1998vh}. For the gauge invariant combinations we have 
computed in our paper, the anomalous dimension matrix given in Ref. \cite{Balzereit:1998jb} yields different RG equations than those we found in 
Sec. \ref{sec:RGeq}, and also different single logs as those given explicitely in that reference. Nevertheless, it is remarkable that 
we find agreement with the explicit single logs given in these references.

These results are necessary building blocks for the determination of the pNRQCD Lagrangian with NNNLL accuracy. Even though 
they appear not to be necessary to obtain the heavy quarkonium spectrum with NNNLL accuracy, nor the production and annihilation of heavy quarkonium 
with NNLL precision, they are necessary at higher orders. Moreover, it may be important for future research in the determination of higher order
logarithms for NRQED bound states, like hydrogen and muonic hydrogenlike atoms.

\medskip

{\bf Acknowledgments} \\
We thank Antonio Pineda for useful discussions and for reading over the manuscript. 
This work was supported in part by the Spanish grants FPA2014-55613-P and SEV-2016-0588. 
RPB was supported for much of the duration of this work by the Barcelona
Institue of Science and Technology (BIST). RPB is currently supported by a Clarendon
Scholarship from the University of Oxford.

\appendix

\section{HQET Feynman rules}
\label{sec:fr}

Here we collect some Feynman rules in Coulomb gauge that are needed for our computation, and complement 
those that can be found in Ref. \cite{Pineda:2011dg} and Ref. \cite{chromopolarizabilities}.  

The heavy quark four-momentum $p=(E_1,\bf p)$ is incoming with associated color index $\beta$, 
whereas the heavy quark four-momentum $p'=(E_1',\bf p\,')$ is outgoing with associated color index $\alpha$. All four-momentums of 
gluons, $ k_i$, are outgoing. If more than one gluon appears, let's say $n$, 
then they are labeled with four-momentum $k_i$ ($i=1,\ldots,n$) and, by four-momentum conservation, $k=\sum_{i=1}^{n}k_i=p-p'$. We start labeling 
transverse gluons first and longitudinal gluons after with the labels $a,b,c,\ldots$ refering to color indices in the adjoint 
representation, $i, j, k,\ldots$ refering to space vector indices and $k_1, k_2, k_3,\ldots$ refering to four-momentum.

\subsection{Proportional to $c_{W_1}$}

\begin{equation}
 \mathcal{V}_{c_{W_1}}^{i\,a} = -c_{W_1}\frac{g}{8m^3}( {\bf p}^2 + {\bf p}'^2)(\bfsigma\times{\bf k})^i (T^a)_{\alpha\beta}
\end{equation}

$$\mathcal{V}_{c_{W_1}}^{ij\,ab}= - c_{W_1}\frac{g^2}{8m^3}
 [(\bfsigma^k \epsilon^{kij}({\bf p}^2 + {\bf p}'^2) - (\bfsigma\times{\bf k}_1)^i {\bf k}_1^j + {\bf k}_2^i(\bfsigma\times {\bf k}_2)^j)[T^a,T^b]_{\alpha\beta}$$
\begin{equation}
 - ((\bfsigma\times {\bf k}_1)^i({\bf p}+{\bf p}')^j + ({\bf p}+{\bf p}')^i(\bfsigma\times{\bf k}_2)^j)\{T^a,T^b\}_{\alpha\beta} ] 
\end{equation}

\begin{equation*}
 \mathcal{V}^{ijk\,abc}_{c_{W_1}}= 
 c_{W_1}\frac{g^3}{8m^3}\bfsigma^m\bigg( \epsilon^{mjk}\big( \{T^a,[T^b,T^c]\}_{\alpha\beta}({\bf p} + {\bf p}')^i 
 - [T^a,[T^b,T^c]]_{\alpha\beta}({\bf k}_2 + {\bf k}_3)^i\big)
\end{equation*}
\begin{equation*}
+ \epsilon^{mki}\big( \{T^b,[T^c,T^a]\}_{\alpha\beta}({\bf p} + {\bf p}')^j
 - [T^b,[T^c,T^a]]_{\alpha\beta}({\bf k}_1 + {\bf k}_3)^j\big)
\end{equation*}
\begin{equation*}
+ \epsilon^{mij}\big( \{T^c,[T^a,T^b]\}_{\alpha\beta}({\bf p} + {\bf p}')^k 
- [T^c,[T^a,T^b]]_{\alpha\beta}({\bf k}_1 + {\bf k}_2)^k \big)
\end{equation*}
\begin{equation*}
-\epsilon^{mri}\delta^{jk} \{T^a,\{T^b,T^c\}\}_{\alpha\beta}{\bf k}_1^r 
- \epsilon^{mrj}\delta^{ik}\{T^b,\{T^c,T^a\}\}_{\alpha\beta}{\bf k}_2^r
\end{equation*}
\begin{equation}
-\epsilon^{mrk}\delta^{ij}\{T^c,\{T^a,T^b\}\}_{\alpha\beta}{\bf k}_3^r\bigg)
\end{equation}

\subsection{Proportional to $c_{W_2}$}

\begin{equation}
 \mathcal{V}_{c_{W_2}}^{i\,a} = c_{W_2}\frac{g}{4m^3}({\bf p}\cdot{\bf p}')(\bfsigma\times{\bf k})^i(T^a)_{\alpha\beta} 
\end{equation}

$$\mathcal{V}^{ij\,ab}_{c_{W_2}}= -c_{W_2}\frac{g^2}{8m^3}[((\bfsigma\times{\bf k}_2)^j {\bf k}_2^i - (\bfsigma\times{\bf k}_1)^i {\bf k}_1^j 
 -2\bfsigma^k \epsilon^{kij}({\bf p}\cdot{\bf p}'))[T^a,T^b]_{\alpha\beta}$$
\begin{equation}
 + ((\bfsigma\times {\bf k}_1)^i ({\bf p}+{\bf p}')^j + ({\bf p}+{\bf p}')^i(\bfsigma\times{\bf k}_2)^j)\{T^a,T^b\}_{\alpha\beta}]
\end{equation}

\begin{equation*}
 \mathcal{V}^{ijk\,abc}_{c_{W_2}}= 
 -c_{W_2}\frac{g^3}{4m^3}\bfsigma^l\bigg( 
 -\epsilon^{lij}([T^a,T^b]T^c)_{\alpha\beta}{\bf k}^k 
 - \epsilon^{lki}([T^c,T^a]T^b)_{\alpha\beta}{\bf k}^j
\end{equation*}
\begin{equation*}
-\epsilon^{ljk}([T^b,T^c]T^a)_{\alpha\beta}{\bf k}^i 
+ \epsilon^{lij}\{T^c,[T^a,T^b]\}_{\alpha\beta} {\bf p}^k 
+ \epsilon^{lki}\{T^b,[T^c,T^a]\}_{\alpha\beta} {\bf p}^j
\end{equation*}
\begin{equation*}
+\epsilon^{ljk}\{T^a,[T^b,T^c]\}_{\alpha\beta} {\bf p}^i 
- \epsilon^{lrj}\delta^{ki}(T^aT^bT^c+T^cT^bT^a)_{\alpha\beta}{\bf k}_2^r
\end{equation*}
\begin{equation}
-\epsilon^{lrk}\delta^{ij}(T^aT^cT^b+T^bT^cT^a)_{\alpha\beta}{\bf k}_3^r 
- \epsilon^{lri}\delta^{jk}(T^bT^aT^c+T^cT^aT^b)_{\alpha\beta}{\bf k}_1^r\bigg)
\end{equation}

\subsection{Proportional to $c_{p'p}$}

\begin{equation}
 \mathcal{V}^{i\,a}_{c_{p'p}} = c_{p'p}\frac{g}{8m^3}\bfsigma\cdot({\bf p}+{\bf p}')({\bf p}\times{\bf p}')^i (T^a)_{\alpha\beta}
\end{equation}

$$\mathcal{V}^{ij\,ab}_{c_{p'p}}= c_{p'p}\frac{g^2}{16m^3}
\bigg\{\big[(({\bf p}+{\bf p}')\times{\bf k}_1)^i \bfsigma^j + \bfsigma^i (({\bf p}+{\bf p}')\times{\bf k}_2)^j 
+ \epsilon^{ijk}({\bf k}_1-{\bf k}_2)^k(\bfsigma\cdot({\bf p}+{\bf p}'))\big]\{T^a,T^b\}_{\alpha\beta} $$
\begin{equation}
 + \big[({\bf k}_1\times{\bf k}_2)^i\bfsigma^j + \bfsigma^i({\bf k}_1\times{\bf k}_2)^j 
 - \epsilon^{ijk}({\bf k}^k(\bfsigma\cdot{\bf k}) + 2((\bfsigma\cdot{\bf p}){\bf p}'^k + (\bfsigma\cdot{\bf p}'){\bf p}^k))\big][T^a,T^b]_{\alpha\beta}\bigg\}
\end{equation}

$$\mathcal{V}^{ijk\,abc}_{c_{p'p}}= c_{p'p}\frac{g^3}{8m^3}
 \bigg[(\epsilon^{ljk}\bfsigma^i {\bf p}^l + \epsilon^{ijk}(\bfsigma\cdot{\bf p}))(T^a[T^b,T^c])_{\alpha\beta} 
 +(\epsilon^{ljk}\bfsigma^i {\bf p}'^l + \epsilon^{ijk}(\bfsigma\cdot{\bf p}'))([T^b,T^c]T^a)_{\alpha\beta}$$
$$+ (\epsilon^{lik}\bfsigma^j {\bf p}^l - \epsilon^{ijk}(\bfsigma\cdot{\bf p}))(T^b[T^a,T^c])_{\alpha\beta} 
 + (\epsilon^{lik}\bfsigma^j {\bf p}'^l - \epsilon^{ijk}(\bfsigma\cdot{\bf p}'))([T^a,T^c]T^b)_{\alpha\beta}$$
 $$+ (\epsilon^{lij}\bfsigma^k {\bf p}^l + \epsilon^{ijk}(\bfsigma\cdot{\bf p}))(T^c[T^a,T^b])_{\alpha\beta}  
 + (\epsilon^{lij}\bfsigma^k {\bf p}'^l + \epsilon^{ijk}(\bfsigma\cdot{\bf p}'))([T^a,T^b]T^c)_{\alpha\beta}$$
$$- (\epsilon^{ljk}\bfsigma^i - \epsilon^{lij}\bfsigma^k){\bf k}_2^l(T^a T^b T^c + T^c T^b T^a)_{\alpha\beta}$$
$$+ (\epsilon^{lik}\bfsigma^j + \epsilon^{ljk}\bfsigma^i){\bf k}_3^l(T^b T^c T^a + T^a T^c T^b)_{\alpha\beta}$$
\begin{equation}
 - (\epsilon^{lij}\bfsigma^k + \epsilon^{lik}\bfsigma^j){\bf k}_1^l(T^c T^a T^b + T^b T^a T^c)_{\alpha\beta}\bigg]
\end{equation}

\subsection{Proportional to $c_{B_1}$}

\begin{equation}
 \mathcal{V}^{ab}_{c_{B_1}}= - c_{B_1}\frac{g^2}{8m^3} \bfsigma \cdot ({\bf k}_1\times {\bf k}_2) [T^a,T^b]_{\alpha \beta}
\end{equation}

\begin{equation}
 \mathcal{V}^{i\,ab}_{c_{B_1}}= - c_{B_1}\frac{g^2}{8m^3} k_1^0 (\bfsigma\times{\bf k}_2)^i [T^a,T^b]_{\alpha \beta}
\end{equation}

$$ \mathcal{V}^{ij\,ab}_{ c_{B_1}}=  c_{B_1}\frac{g^2}{16m^3}
 \bigg( \epsilon^{ijk}(\bfsigma\cdot{\bf k}_1){\bf k}_2^k + \epsilon^{ijk}(\bfsigma\cdot{\bf k}_2){\bf k}_1^k 
 + \bfsigma^i({\bf k}_1\times{\bf k}_2)^j + \bfsigma^j({\bf k}_1\times{\bf k}_2)^i$$
\begin{equation}
 - 2\bfsigma^k \epsilon^{kij} k^0_1 k^0_2\bigg)[T^a,T^b]_{\alpha\beta}
\end{equation}

\begin{equation}
 \mathcal{V}^{i\,abc}_{ c_{B_1}}= - c_{B_1} \frac{g^3}{8m^3} 
 ( ( \bfsigma\times{\bf k}_2)^i [T^b,[T^a,T^c]]_{\alpha \beta} +  (\bfsigma\times{\bf k}_3)^i [T^c,[T^a,T^b]]_{\alpha \beta}) 
\end{equation}

\begin{equation}
 \mathcal{V}^{ij\,abc}_{c_{B_1}} =  c_{B_1}\frac{g^3}{8m^3} \bfsigma^k \epsilon^{kij} 
 (   k_1^0 [T^a,[T^b,T^c]]_{\alpha \beta} - k_2^0 [T^b,[T^a,T^c]]_{\alpha \beta} ) 
\end{equation}

\begin{align}
 \mathcal{V}^{ijk\,abc}_{c_{B_1}}=&   - c_{B_1}\frac{g^3}{8m^3} 
 \left\{ \left(  \epsilon^{ijk} \bfsigma \cdot {\bf k}_1 - \epsilon^{ljk} \bfsigma^i  {\bf k}^l_1  \right) [T^a,[T^b,T^c]]_{\alpha\beta} \right. \nonumber\\
 &-\left( \epsilon^{ijk} \bfsigma \cdot {\bf k}_2 -  \epsilon^{ilk}\bfsigma^j {\bf k}^l_2  \right) [T^b,[T^a,T^c]]_{\alpha\beta}  \nonumber \\
 &\left.+\left(  \epsilon^{ijk} \bfsigma \cdot {\bf k}_3 - \epsilon^{lij} \bfsigma^k  {\bf k}^l_3  \right) [T^c,[T^a,T^b]]_{\alpha\beta}  \right\}
\end{align}

\subsection{Proportional to $c_{B_2}$}

\begin{equation}
 \mathcal{V}^{ab}_{c_{B_2}}= - c_{B_2}\frac{g^2}{8m^3} \bfsigma \cdot ({\bf k}_1\times {\bf k}_2) [T^a,T^b]_{\alpha \beta}
\end{equation}

\begin{equation}
 \mathcal{V}^{i\,ab}_{c_{B_2}}= - c_{B_2}\frac{g^2}{8m^3} k_1^0 (\bfsigma\times{\bf k}_2)^i [T^a,T^b]_{\alpha \beta}
\end{equation}

\begin{equation}
 \mathcal{V}^{ij\,ab}_{c_{B_2}}= - c_{B_2}\frac{g^2}{8m^3} \bfsigma^k \epsilon^{kij} k^0_1 k^0_2[T^a,T^b]_{\alpha \beta}
\end{equation}

\begin{equation}
 \mathcal{V}^{i\,abc}_{c_{B_2}}= - c_{B_2} \frac{g^3}{8m^3} 
 ( ( \bfsigma\times{\bf k}_2)^i [T^b,[T^a,T^c]]_{\alpha \beta} +  (\bfsigma\times{\bf k}_3)^i [T^c,[T^a,T^b]]_{\alpha \beta}) 
\end{equation}

\begin{equation}
 \mathcal{V}^{ij\,abc}_{c_{B_2}} = c_{B_2}\frac{g^3}{8m^3} \bfsigma^k \epsilon^{kij} 
 (   k_1^0 [T^a,[T^b,T^c]]_{\alpha \beta} - k_2^0 [T^b,[T^a,T^c]]_{\alpha \beta} ) 
\end{equation}

%%%%%%%%%%%%%%%%%%%%% BIBLIOGRAPHY %%%%%%%%%%%%%%%%%%%%%%%%%%%%%%%%%%%%

\end{document}